# Efficient Design Scheme of Superconducting Cavity

Sang-ho Kim, Marc Doleans, SNS/ORNL, USA
Yoon Kang, APS/ANL, USA

*Abstract*

For many next-generation high intensity proton accelerator applications including the Spallation Neutron Source (SNS), superconducting (SC) RF provides the technology of choice for the linac. In designing the superconducting cavity, several features, such as peak fields, inter-cell coupling, mechanical stiffness, field flatness, external Q, manufacturability, shunt impedance, higher order mode (HOM), etc., should be considered together. A systematic approach to determine the optimum cavity shape by exploring the entire geometric space of the cavity has been found. The most efficient use of RF energy can be accomplished by adjusting the cell shape. A small region in parameter space satisfying all reasonable design criteria has been found. With this design procedure, choosing the optimum shape is simplified. In this paper, the whole design procedure of this optimisation scheme is explained and applied to the SNS cavity design.

## 1 INTRODUCTION

In many recently initiated or proposed projects for high intensity proton acceleration, SCRF technology has been selected for the main part of the linac, which uses elliptical shape SC cavities. SNS will be the first high intensity proton accelerator with a SC linac. The basic parameters of the SNS SC linac are shown in Table 1.

Table 1: Basic parameters of the SNS SC linac

| | |
|---|---|
| RF frequency | 805 MHz |
| Energy range | 185-1000 MeV |
| Average beam current | 2 mA |
| Number of beta sections | 2 (0.61 and 0.81) |
| Transition energy between sections | ~380 MeV |
| Cavity shape | elliptical (6 cells) |

In designing the cavity, RF and mechanical properties are considered together, especially for the cavity whose beta is less than one. The general design bases and issues for the SNS cavity are summarized in terms of cavity parameters.

For the inner cell design;

- Minimise the peak surface fields
- Provide a reasonable mechanical stiffness
- Maximize the R/Q
- Achieve a reasonable inter-cell coupling coefficient

For the end cell and full cavity design;

- Obtain a good field flatness
- Obtain a lower (or same) surface fields at end cells than (or with) those of inner cell
- Achieve a reasonable external Q, Qex

All the issues listed above are directly linked to the shape, and the effects of shape on these issues are different. In some aspects, the effects compete, and optimization among tradeoffs becomes necessary. A systematic scheme is introduced here for choosing the optimum cavity shape.

## 2 INNER-CELL DESIGN

Figure 1 shows the geometric parameters of the elliptical cell. Adjusting four of these five parameters (Req, α, Rc, a/b, Ri) determines a cell shape that satisfies required beta and frequency. Usually the equator radius is used for tuning, since its effect on the resonance frequencies is large and its influence on the other cavity parameters is negligible.

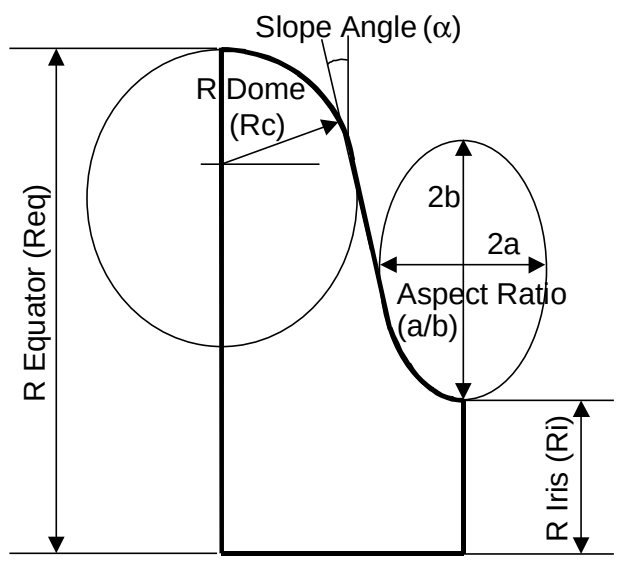

Figure 1: Geometric parameters of the cell.

In order to understand the influences of cell parameters on the cavity performance, the entire geometric space was explored. The following procedures were established from this understanding. Peak fields (Ep & Bp), inter-cell coupling coefficient (k), R/Q, and Lorentz force detuning coefficient (K) are used as cavity parameters. The first step is to determine relations between the dome radius and the iris ellipse aspect ratio at fixed iris radius and slope angle. At any dome radius, the surface electric field profile can be changed by adjusting a/b. In this adjustment variations of other cavity parameters are negligible. Figures 2 (a) and (b) are comparisons of surface electric fields for given a peak surface electric field and accelerating field, respectively. The line 2 in Figure 2 (a) has higher accelerating field than the others. The line 2 in Figure 2 (b) has lower peak electric field than the others. The field profile of line 2 provides the efficient use of RF energy.

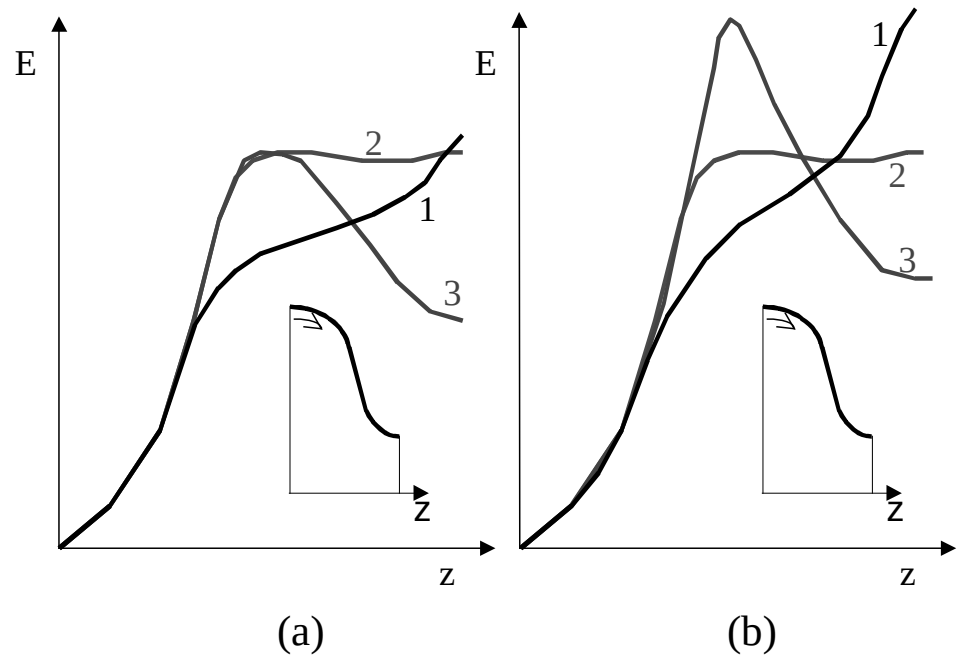

Figure 2: Surface electric fields of cells with different values of a/b; (a) at same peak surface electric field, (b) at same acceleration gradient.

The best values of a/b are automatically determined by fixing the other geometric cell parameters. In this procedure, a/b's are found as a function of Rc's at given Ri and α. Slope angles above 6 degree are required for the rinsing process. Due to its small cell length, only a small angle region from 6 to 8 degrees leads to a good cell performance for the medium beta case. Finally the cell geometry can be defined with the remaining two cell parameters, Ri and Rc, at fixed slope angle.

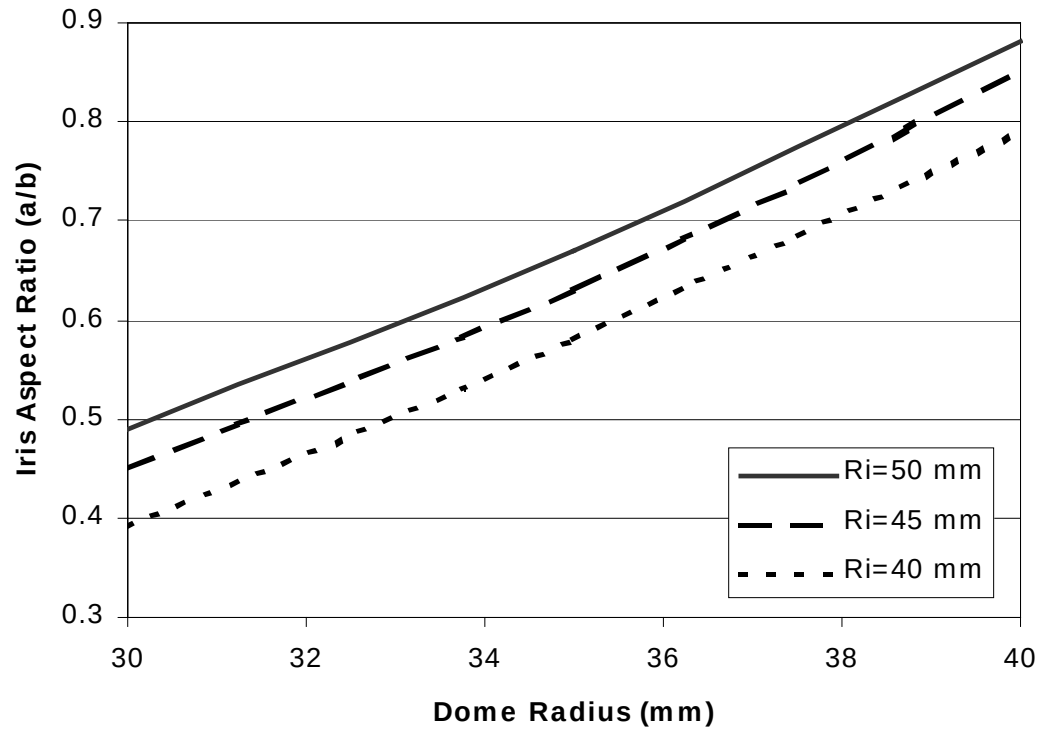

Figure 3: "Efficient-set" lines of cell geometry for SNS b=0.61 cavity at the slope angle of 7 degree.

Figure 3 is an example of SNS medium beta (β=0.61) inner cell at the slope angle of 7 degrees. SUPERFISH was used for the analysis [1]. These lines all satisfy the condition of a flat field around the iris. In Figure 4, relative values of cavity parameters are plotted for the cell geometry on the solid line in Figure 3. The cell with efficient-set having Rc=30 mm is used as normalisation reference. The Lorentz force detuning coefficient K is calculated with fixed boundary condition and stiffener location of 70 mm from the cavity axis. This coefficient is sensitive to the shape of the iris ellipse, especially in low beta case. Similar graphs can be done for other Ri values to cover all the geometric parameter space.

An optimum cell shape that satisfies all the design criteria can be found by combining the results from different Ri's.

Figure 5 is an example of SNS medium beta cavity at the slope angle of 7 degree. The SNS design criteria are; Ep=27.5 MV/m, Bp<60 mT, k>1.5 %, K<3 Hz/(MV/m)$^2$, and Eo>11.9 MV/m for the reference geometry.

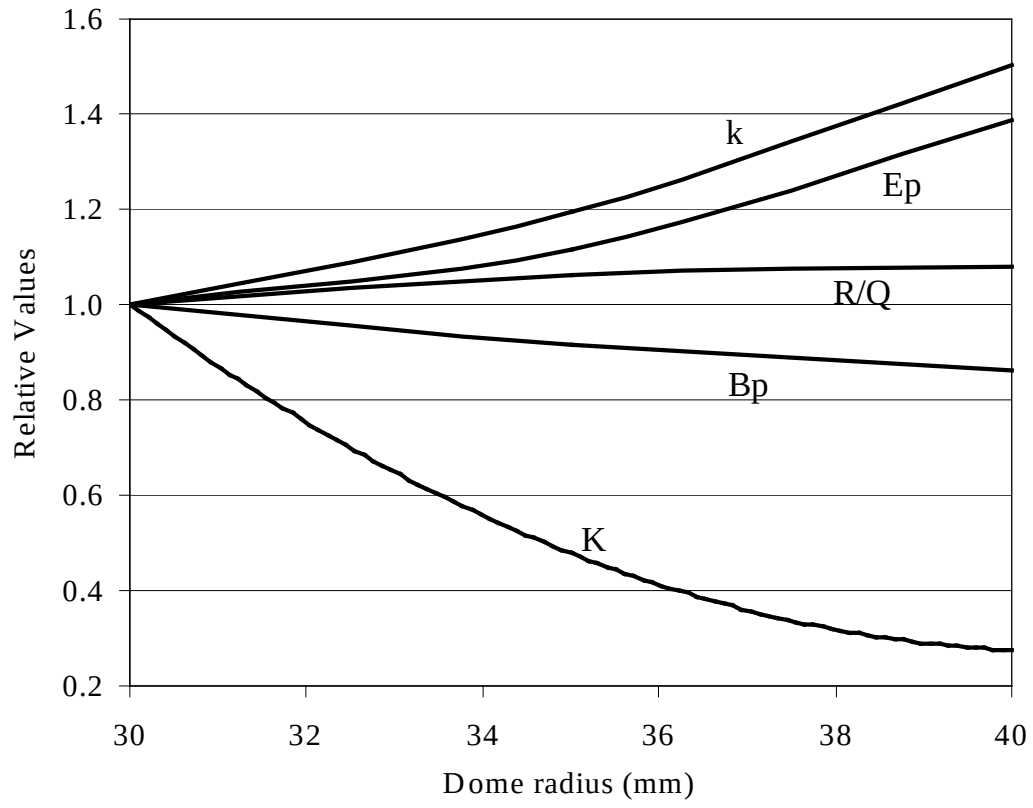

Figure 4: Relative cavity parameter behaviors for Ri=50 mm and a=7 degree versus Rc. "Efficient-set" geometry is represented by dome radius, Rc.

The Eo value used here pertains to the inner-cell only. All the design criteria are marked with bold lines in Figure 5. There is a small region where all the design criteria are satisfied. Selecting the final cell geometry is a matter of the strategy. SNS chose the cell geometry for the high accelerating gradient within the design criteria.

The results of inner cell design for the high beta cavity show similar behaviours except a few aspects. The Lorentz force detuning is not sensitive to cell shape, so this is not an issue in the high beta cavity. The slope angle can be chosen from 6 to 12 degrees, for specified cavity parameters. The larger slope angle is better for end cell tuning.

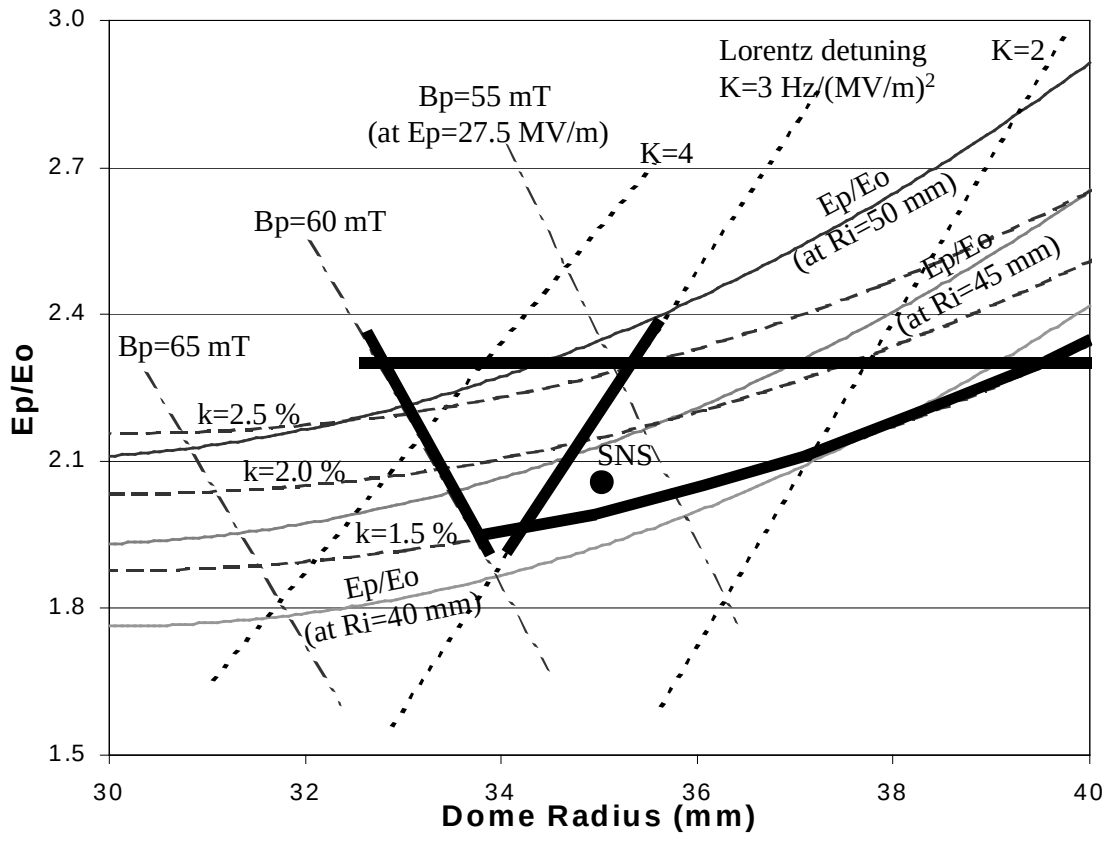

Figure 5: Overall comparisons of cavity parameters on the cell geometric parameter.

## 3 END CELL DESIGN

End-cells should be tuned separately due to the attached beam pipes. Changing the shape of end-cells must lead to a reasonable axial electrical field flatness below ~2 %. Peak surface fields must be equal or lower than inner cells value. Many different end-cell shapes can satisfy these criteria. The Figure 6 shows the axial electric field profile for different acceptable cavity geometries.

Each end-cell designs necessitate a different approach because one is connected to the power coupler. A coaxial type power coupler will be used in SNS.

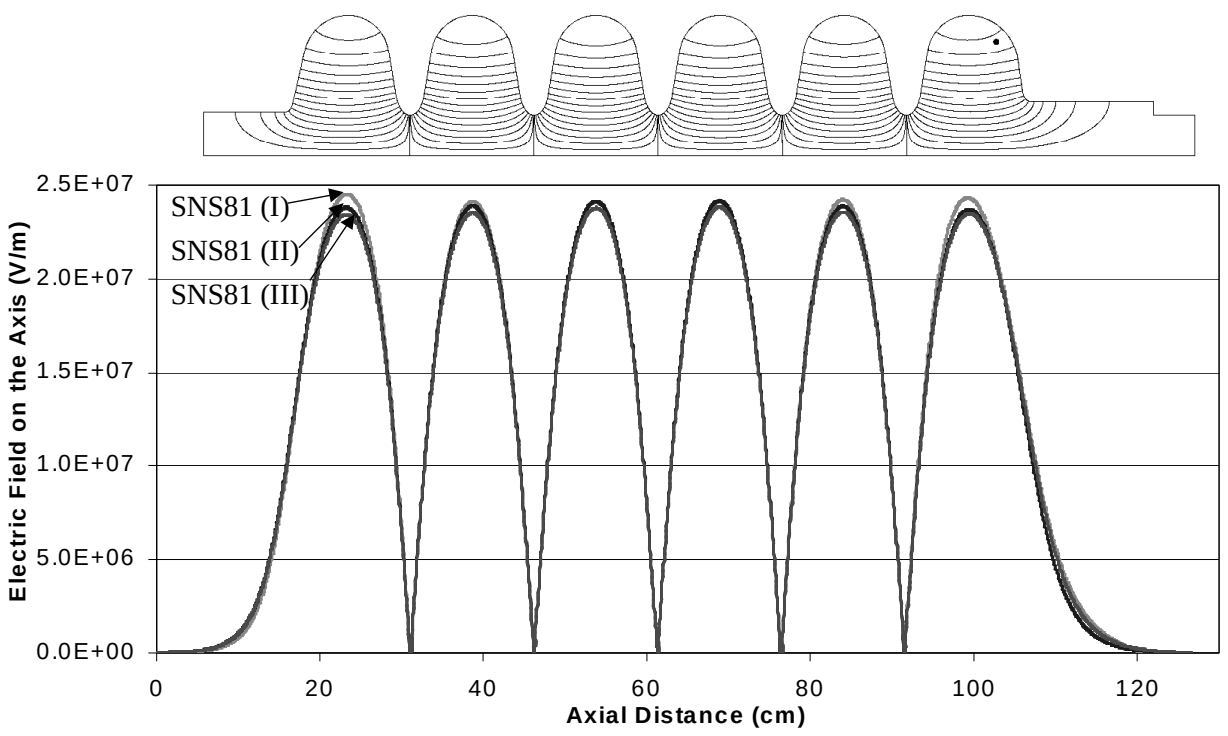

Figure 6: Axial electric field profiles for three different SNS high beta ($\beta$=0.81) cavities at Ep=27.5 MV/m.

Required Qex's are $7.3\times10^5$ and $7.0\times10^5$ for medium and high beta cavities, respectively. A computer study of Qex has been done following a scheme introduced in [2]. Four-parameter space has been explored (Figure 7). The geometry of the power coupler is not used as a parameter in the study.

The inner conductor tip position has a strong effect on the coupling between the cavity and the power coupler. As shown in Figure 8, about 25 mm displacement results in one order variation on the Qex value.

1) GOC
(Geometry of Coupler)
2) BPR
(Beam Pipe Radius)
3) REG
(Right End-cell Geometry)
4) LPC
(Longitudinal Position of Coupler)
5) ITP
(Inner conductor Tip Position)

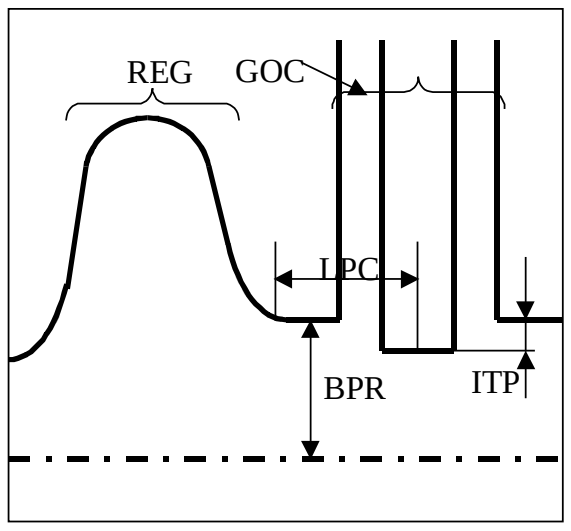

Figure 7: Five parameters that can affect Qex.

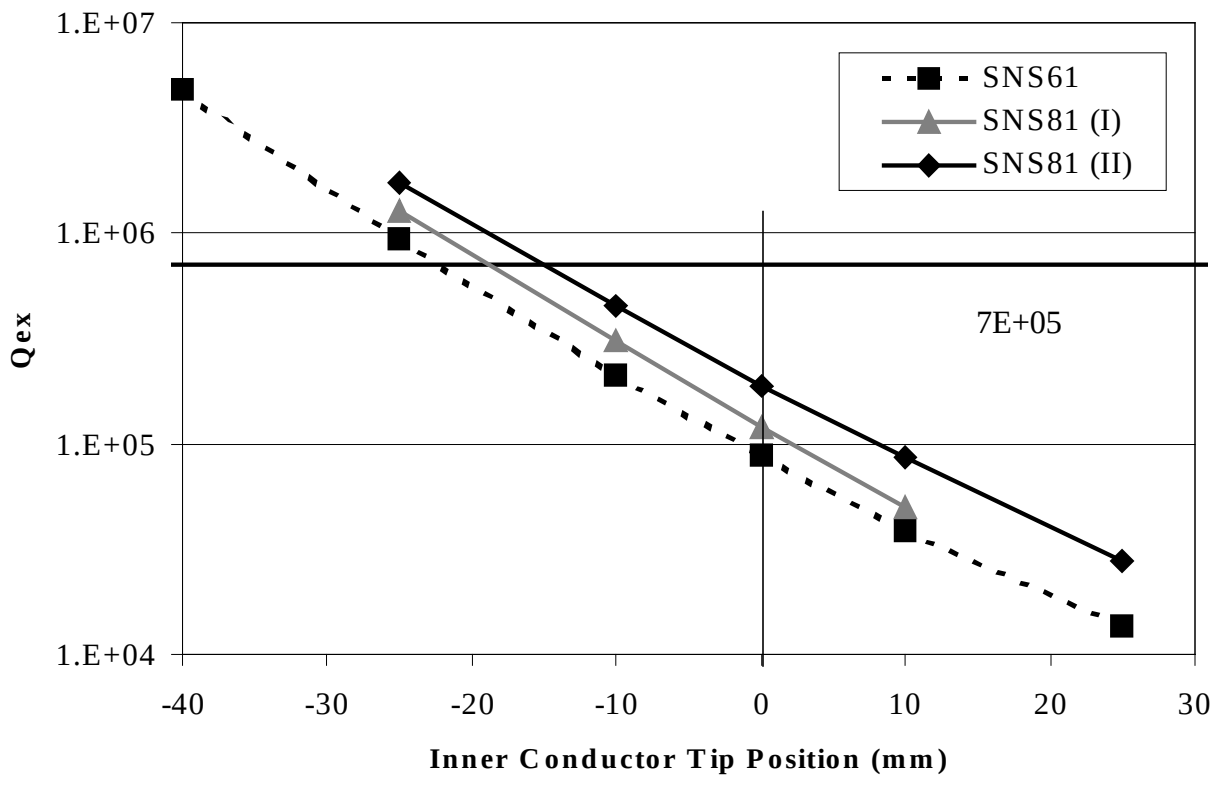

Figure 8: Variations of Qex's as a function of ITP for three different cavities. Same GOC and 7 cm of LPC are used for each calculation.

Since keeping the same iris radius as the inner cell leads to a high Qex value, enlarging the beam pipe size can provide a solution. This option is efficient below a certain diameter. Figure 9 shows that over 62 mm, increasing the beam pipe size has a weak influence on the high beta case. The points marked with triangle are not on the line. This results from the change in field profile after end cell tuning. These points still satisfy all requirements, that means the Qex can be also controlled by changing end cell shape only.

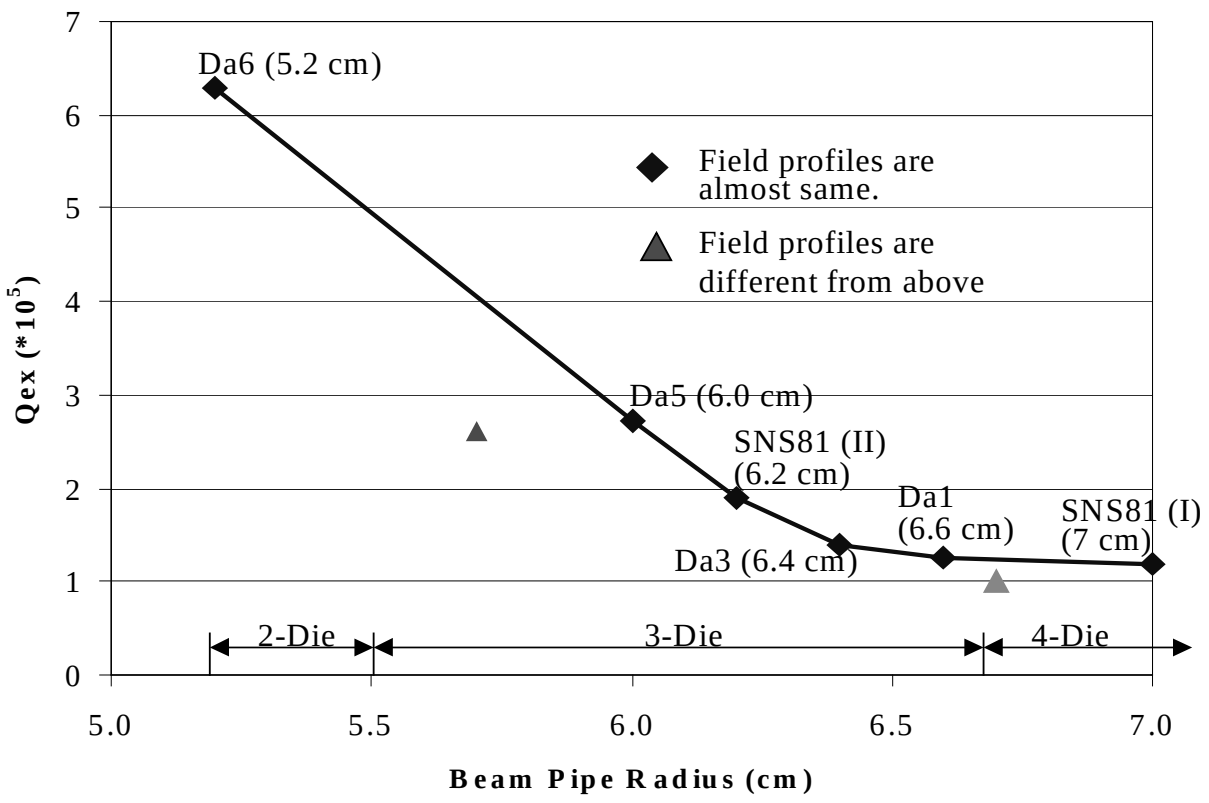

Figure 9: Variation of Qex's as a function of BPR for SNS high beta cavity. Same GOC, 0 cm of ITP and 7 cm of LPC are used for each calculation.

The effect of LPC is also examined from 8.5 cm to 7 cm, the lowest possible distance. The Qex decreases linearly by a factor of three.

Many end cell shapes could satisfy the requirements. The final decision will depend on the amount of engineering margin.

## 4 HIGHER ORDER MODE (HOM)

HOM analysis for the reference geometry has been done. Many trapped modes are found even in reference geometry. Beam dynamics issues related with HOM are under study and the intermediate results suggest that the cumulative beam break-up is not an issue in SNS [3]. In order to investigate the effects of mechanical imperfection of the cavity on the trapped modes, Monte-Carlo analysis is in progress.

## 5 SUMMARY

The cavity performance is visualised in the geometric parameter space by the systematic scheme introduced.

## 6 ACKNOWLEDGEMENT

We are grateful to R. Sundelin, P. Kneisel at Jlab, and James Billen at LANL for giving many useful advises and comments.

This work is sponsored by the Division of Materials Science, U.S.Department of Energy, under contraction number DE-AC05-96OR22464 with UT-Bettelle Corporation for Oak Ridge National Laboratory.